\documentstyle[sprocl,epsfig]{article}

\newcommand{\inttp}{\int \frac{d^3p}{(2 \pi)^3}}

\newcommand{\ihsp}{\hspace*{\fill} }

\newcommand{\case}[2]{\mbox{\small $\displaystyle \frac{#1}{#2}$}}

\newcommand{\bcn}{\begin{center}}
\newcommand{\beq}{\begin{equation}}
\newcommand{\beqar}{\begin{eqnarray}}

\newcommand{\ecn}{\end{center}}
\newcommand{\eeq}{\end{equation}}
\newcommand{\eeqar}{\end{eqnarray}}

\newcommand{\Eq}[1]{Eq.~(\ref{#1})}
\newcommand{\Eqs}[1]{Eqs.~(\ref{#1})}
\newcommand{\Fig}[1]{Fig.~\ref{#1}}

\newcommand{\Table}[1]{Table~\ref{#1}}
\newcommand{\sect}[1]{\section{ #1} }
 
\def\be{\begin{equation}}
\def\ee{\end{equation}}
\def\bea{\begin{eqnarray}}
\def\eea{\end{eqnarray}}
\begin{document}

\parbox{108mm}{ 
%\hspace*{\fill}
\footnotesize
Contribution to the IVth Workshop on Quantum Chromodynamics,  
1-6 June, 1998, Paris. Proceedings to be published. \hspace*{\fill}
\parbox[t]{25mm}{KSUCNR-107-98 \\ }  }
\normalsize
%\vspace*{-0.5cm}

\title{MODELING NONPERTURBATIVE QCD FOR MESONS AND COUPLINGS}

\author{PETER C. TANDY}

\address{Center for Nuclear Research, Department of Physics\\
Kent State University, Kent, OH, USA 44242\\E-mail: tandy@cnr2.kent.edu} 


\maketitle\abstracts{ We discuss aspects of a covariant QCD
modeling of meson physics by illustrating applications to several 
coupling constants and form factors.  In particular, we cover the 
$\rho\pi\pi$ and $\pi^0 \gamma \gamma$ interactions, the $\rho$ 
contribution to the pion charge radius, and $\pi NN$ coupling.  }

\section{Introduction}

The  Dyson-Schwinger equation (DSE) approach\cite{DSErev} to 
non-perturbative QCD modeling provides semi-phenomenological gluon and 
quark two-point functions that have proved to be quite efficient in 
describing and predicting the physics of low-mass mesons \cite{T97}.
To outline the basis of such efforts, consider 
the fully-dressed and renormalized quark propagator defined by 
the  Dyson-Schwinger equation (DSE) in Euclidean metric
\begin{equation}
 S^{-1}(p) = Z_2 [i \gamma\cdot p + m_0(\Lambda)]
     + Z_1 \; \case{4}{3} \int^\Lambda \frac{d^4k}{(2\pi)^4} g^2 D_{\mu \nu}(p-k)
        \gamma_\mu S(k) \Gamma_\nu^g (k,p) ,     \label{fullDSE}
\end{equation}
where $m_0(\Lambda)$ is the bare mass parameter and $\Lambda$ 
characterizes the regularization mass scale. Here the dressed gluon 
propagator
$D_{\mu\nu}(q)$  and the dressed quark-gluon vertex $\Gamma_\mu^g(k,p)$ 
are the renormalized quantities and they satisfy their own DSEs which
require at least some higher order $n$-point functions.  The 
ultraviolet behavior of $D_{\mu\nu}(q)$ and $\Gamma_\mu^g(k,p)$ can be
constrained by perturbation theory.  However physics at the hadron length
scale depends crucially upon the behavior of $S(p)$, $D_{\mu\nu}(q)$ and 
$\Gamma_\mu^g(k,p)$ in the infrared. Present QCD modeling begins 
with an explicit solution of the quark DSE via phenomenological IR forms for
$D_{\mu\nu}(q)$ and $\Gamma_\mu^g(k,p)$.  Although the DSE itself can 
probe the pseudoscalar component of the pion via the chiral condensate 
and dynamical chiral symmetry breaking, the Bethe-Salpeter equation (BSE) 
is required to address other mesons and to obtain the sub-leading pion
components. 

The BSE  for a bound state of a quark of
flavor $f_1$ and an antiquark of flavor $\bar f_2$ is
\begin{equation}
\Gamma(p;P) =  \int \frac{d^4q}{(2\pi)^4} K(p,q;P) S_{f_1}(q + \xi P) 
                     \Gamma(q;P) S_{f_2}(q - \bar\xi P)   ~, \label{bspi}
\end{equation}
where \mbox{$\xi + \bar\xi = 1$} describes momentum sharing. 
The kernel $K$ operates in the direct product space of color, flavor and
Dirac spin for the quark and antiquark and 
is the renormalized, amputated $\bar q q$ scattering
kernel that is irreducible with 
respect to a pair of $\bar q q$ lines. 
At the present stage of QCD modeling, the BSE is employed in ladder 
approximation with bare vertices, that is
\beq
K(p,q;P) = -g^2 D_{\mu \nu}(p-q) \case{\lambda^a}{2} \gamma_\mu \otimes
 \case{\lambda^a}{2} \gamma_\nu ~.
\label{ladderK}
\eeq
The treatment of the quark DSE that is dynamically matched to this is
the bare vertex or rainbow approximation for then the axial vector 
Ward-Takahashi identity is preserved and the Goldstone theorem is manifest.
The various models of this type generally use the Ansatz
\begin{equation}
\label{Deff}
g^2 D_{\mu\nu}(q) \rightarrow
        \left(\delta_{\mu\nu} - \frac{q_\mu q_\nu}{q^2}\right)
                \frac{4\pi\alpha_{\hbox{\rm eff}}(q^2)}{q^2}~,
\end{equation}
where $\alpha_{\hbox{\rm eff}}(q^2)$ implements the pQCD running coupling in
the UV and a phenomenological enhancement in the IR.  The pion and kaon are 
well described in a recent work of this type.~\cite{MR97}

\section{QCD Modeling of Mesons and Interactions}

In Euclidean metric, the mass-shell condition for meson couplings requires
the quark propagators in loop calculations be evaluated at complex quark
momentum.  To facilitate a broad survey of such applications, the
present approach is to  make use of a convenient analytic parameterization 
of confined solutions of the quark DSE.  The broad features  are taken
from the solution to a simple DSE model~\cite{BRW92} that is extremely 
infrared dominant, produces a propagator with no mass-shell pole, and 
includes gluon-quark vertex dressing determined by the Ward
identity.  The resulting propagator is an entire function in 
the complex $p^2$-plane describing absolutely confined~\cite{RWK92} 
dressed quarks in the presence of both explicit and dynamical breaking 
of chiral symmetry.   The finer details of more realistic DSE solutions are
accommodated by typically five parameters that are are used to restore 
a good description of  pion and kaon observables:
$f_{\pi/K}$; $m_{\pi/K}$; $\langle\bar q q\rangle$; $r_\pi$; the $\pi$-$\pi$ 
scattering lengths; and the electromagnetic pion form 
factor.~\cite{R96,BRT96}

\begin{figure}[ht]
\ihsp \epsfig{figure=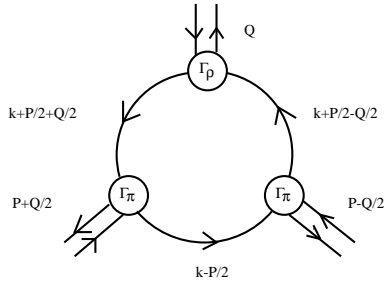,height=5.0cm} \ihsp
\caption{Diagram for the $\rho\pi\pi$ calculation.
\label{diag:rpp} } 
\end{figure}

The general form of the pion Bethe-Salpeter (BS) amplitude is
\begin{eqnarray}
\Gamma_\pi^j(k;P) & = &  \tau^j \gamma_5 \left[ i E_\pi(k;P) + 
\gamma\cdot P F_\pi(k;P) \rule{0mm}{5mm}\right. \nonumber \\
& & \left. \rule{0mm}{5mm}+ \gamma\cdot k \,k \cdot P\, G_\pi(k;P) 
+ \sigma_{\mu\nu}\,k_\mu P_\nu \,H_\pi(k;P) \right]\, ,
\label{Gammapi}
\end{eqnarray}
and the first three terms are significant in realistic model 
solutions~\cite{MR97} and are necessary to satisfy the axial
Ward-Takahashi identity.~\cite{MRT98}   The latter identity, to lowest order
in $P$ at the chiral limit, yields \mbox{$E_\pi(k;P=0) = B(k^2)/f_\pi$}
where $B$ is the scalar part of the quark dynamical 
self-energy.~\cite{MRT98}   It has been quite common to assume that only 
this term of $\Gamma_\pi$ is important for pion coupling.  This has been
questioned by recent studies of interactions, such as that shown in  
Fig.~\ref{diag:rpp}, where we employ approximate
$\pi$ BS amplitudes such as those obtained from a rank-2 separable 
ansatz~\cite{sep97} for the ladder/rainbow kernel of the DSE and BSE.  
They preserve Goldstone's theorem and should be adequate for infrared
integrated quantities.   Parameters are fit to $m_{\pi/K}$ and 
$f_{\pi/K}$. The resulting $\pi$ BS amplitude is
\beq
\Gamma_\pi (k,Q) = i\gamma_5 f(k^2)\, \lambda_1^\pi\, -
      \gamma_5 \,\gamma\cdot Q f(k^2)\, \lambda_2^\pi~. 
\label{sep_pi}
\eeq
The transverse amplitude for the $\rho$ from the same study~\cite{sep97} is
\beq
\Gamma_\nu^\rho(k;Q) = k_{\nu}^Tg(k^2)\lambda_1^\rho 
+ i\gamma_{\nu}^T f(k^2)\lambda_2^\rho 
+ i\gamma_5 \epsilon_{\mu \nu \lambda \rho}
\gamma_{\mu} k_{\lambda} Q_{\rho}  g(k^2)\lambda_3^\rho .
\label{vamp}
\eeq
The BS amplitudes are normalized in the canonical way. 
%----------rho pi pi diagram----------------------------------- 
%-------------------------------------------------------------

\sect{The $\rho\pi\pi$, $\pi^0 \gamma \gamma$ and $\gamma \pi \pi$ 
            Interactions}
\label{sect_rpp}

The first term in a skeleton graph expansion of the $\rho\pi\pi$ 
vertex~\cite{MT97} can be expressed as
\beq
\Lambda_\mu(P,Q)= 2 N_c {\rm tr}_s \int \frac{d^4k}{(2\pi)^4} 
        S(q^\prime)\Gamma^\rho_\mu S(q^{\prime\prime}) 
         \Gamma_\pi S(q^{\prime\prime\prime}) \Gamma_\pi ~,
\label{vint}
\eeq
The momentum notation can easily be deduced from the Feynman rules associated
with Fig.~\ref{diag:rpp}, e.g. \mbox{$q^\prime= k+P/2+Q/2$} and
$\Gamma^\rho_\mu$ represents $\Gamma^\rho_\mu(k+P/2;Q)$ etc.  
With both pions on the mass-shell,
$P\cdot Q=0$ and \mbox{$P^2=$}\mbox{$-m_\pi^2-Q^2/4$}.  In this
case symmetries require the form
\mbox{$\Lambda_\mu(P,Q)=$}\mbox{$-P_\mu  F_{\rho\pi\pi}(Q^2)$} and
the coupling constant is 
\mbox{$ g_{\rho\pi\pi}=$}\mbox{$ F_{\rho\pi\pi}(Q^2=-m_\rho^2)$}.  

Previous investigations of the $\rho\pi\pi$ coupling constant in terms
of a covariant quark-gluon phenomenology for the intrinsic properties
of $\rho$ and $\pi$ employed only $\gamma_\mu$ and $\gamma_5$ covariants
for the respective BS amplitudes.~\cite{MT97,PRC87bHRM92}  
It has since been demonstrated for a number of infra-red sensitive 
quantities such as $m_\pi$ and $f_\pi$, that the pseudovector terms
in the pion BS amplitude are responsible for corrections
in the 20-30\% range.~\cite{MR97,MRT98,sep97}   The model $\rho$ amplitude
in \Eq{vamp} is admittedly crude, but the relative magnitude of the three
surviving scalar amplitudes will hopefuly provide qualitative guidance.  
%---------rho-pi-pi result Table------------------------------
\begin{table}[ht]
\caption{$g_{\rho\pi\pi}$ calculation and contributions from meson 
covariants. \label{tab:rpp} }
\vspace{0.2cm}
\begin{center}
\footnotesize
\begin{tabular}{|c|}\hline
 $g_{\rho\pi\pi}= 6.28$  [expt 6.05] \\
\end{tabular}\\
\begin{tabular}{|cl||cl|} \hline
  $\pi$ Covariants  &    & $\rho$ Covariants & \\ \hline
 $\gamma_5$ & 171\%  & $\gamma_\mu$ & 94.5\%   \\ 
 $\gamma_5 \gamma \cdot Q$ & -71\% & $\gamma_5 \epsilon_\mu \; 
                                           \gamma k Q$ & 5.5\% \\ 
            &      &     $k_\mu$   &     0.01\%  \\ \hline
\end{tabular}
\end{center}
\end{table}
%---------------------------------------
With the separable model BS amplitudes of \Eqs{sep_pi} and (\ref{vamp}),
the prediction for $g_{\rho\pi\pi}$, given in \Table{tab:rpp}, compares
favorably with the empirical value associated with the 
$\rho\rightarrow\pi\pi$ decay width. Truncation to the dominant 
$\rho$ amplitude is found
to only make a 5\% error.  However the sub-dominant pion component 
(pseudovector \mbox{$\gamma_5 \gamma \cdot Q$}) enters quadratically here
and makes a major contribution (-71\%). 

Studies of pion loops  in the $\rho-\omega$ 
sector~\cite{MT97,PRC87bHRM92} and in the pion charge form 
factor~\cite{ABR95} suggest that the $\bar q q$ extended 
structure of the pion significantly weakens such contributions 
compared to models or effective field theories built on point coupling. 
A similar issue arises in the role of the $\rho$ in the 
space-like pion charge form factor.   The dressed photon-quark 
vertex $\Gamma_\nu(q;Q)$ can be separated (non-uniquely) into a 
$\rho$ pole or resonant piece (which is transverse) and a background or 
non-resonant piece (which is both longitudinal and transverse).  Thus 
in the present approach the pion charge form factor takes the form
\beq
F_\pi(Q^2) = F_\pi^{GIA}(Q^2) + \frac{ F_{\rho\pi\pi}(Q^2) 
                   \; \Pi^{\rho \gamma}_T(Q^2) } {Q^2 + m_\rho^2(Q^2)} ~,
\label{piffrho}
\eeq
where $\Pi^{\rho \gamma}_T(Q^2)$ is the $\rho \gamma$ polarization tensor
and  $F_\pi^{GIA}(Q^2)$ is the 
generalized impulse approximation (GIA) result due to the non-resonant
photon-quark coupling.  It has been found to be 
phenomenologically successful in the spacelike region and 
a persistent result  is that $85-90$\% of the charge radius 
is naturally explained that way.~\cite{R96}  

Is the $\rho$ contribution small enough in the present QCD-modeling 
approach?  With
\mbox{$F_{\rho \pi\pi}(Q^2) =  g_{\rho \pi\pi} f_{\rho \pi\pi}(Q^2)$} and
\mbox{$\Pi^{\rho \gamma}_T(Q^2) = -Q^2 f_{\rho \gamma}(Q^2)/g_V $}, 
which is consistent with electromagnetic gauge invariance, 
both form factors $f$  depend on meson substructure dynamics and 
have been calculated.   
The $\rho$ contribution to $r_\pi$ from \Eq{piffrho} is then
\beq
(r_\pi^{pole})^2 = r_\pi^2 - (r_\pi^{GIA})^2 = 1.2 \;
 f_{\rho \pi\pi}(0)  f_{\rho \gamma}(0) \; \frac{6}{m_\rho^2}  ~,
\label{rpipole}
\eeq
where we have used the empirical result 
\mbox{$ g_{\rho \pi\pi}/g_V \sim 1.2$} rather than universal vector coupling.
Our calculations include the extended nature of the mesons and produce 
\mbox{$f_{\rho \pi\pi}(0) \approx$} 
\mbox{$ 0.5$} and \mbox{$ f_{\rho \gamma}(0) \approx 0.65$}.  This 
yields  \mbox{$(r_\pi^{pole})^2 =$} \mbox{$ 0.16~{\rm fm}^2$}.  
In contrast, the empirical Vector Meson Dominance 
(VMD) picture has \mbox{$r_\pi^{GIA}=0$} and \mbox{$f_{\rho \pi \pi}=$}
\mbox{$f_{\rho \gamma}=1$} so that
\mbox{$r_\pi^2 \sim 6  g_{\rho \pi\pi}/(m_\rho^2 g_V) $}
and produces $\sim 0.4~{\rm fm}^2$.  Adding the non-resonant impulse 
result~\cite{R96} \mbox{$(r_\pi^{GIA})^2 =$} \mbox{$ 0.31~{\rm fm}^2$}
gives a total of $0.47~{\rm fm}^2$ with our present approach.  This is 
obviously an
overestimate of the experimental value ($0.44~{\rm fm}^2$) leaving no room 
for the pion loop contribution of the expected~\cite{ABR95} size.
However, the main point is that a $\rho$ contribution to the pion charge
radius which is much smaller than that
from the simple VMD assumption is consistent with the present status of
DSE-based QCD modeling of the pion.

%----------pi gamma gamma figure----------------------------------- 
\begin{figure}[ht]
\ihsp \epsfig{figure=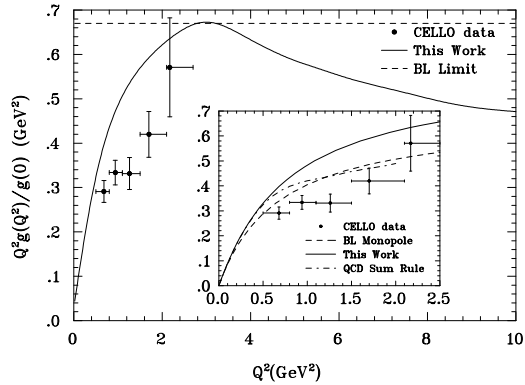,angle=90,height=5.0cm} \ihsp
\caption{The $\pi^0 \gamma \gamma $ transition form factor.
\label{fig:pgg} } 
\end{figure}
%-------------------------------------------------------------

The coupling constant for the \mbox{$\pi^0 \rightarrow \gamma \gamma$} decay
is given by the axial anomaly and is a consequence of gauge invariance and
chiral symmetry in quantum field theory.   A modeling of nonperturbative QCD
should preserve these features.  This was the aim of a study~\cite{FMRT95}
that applied the present approach and also investigated the form factor for 
the transition \mbox{$\gamma^\ast \pi^0 \rightarrow \gamma$}.  The
calculation employed a dressed quark loop similar to Fig.~\ref{diag:rpp}.
The dressed photon-quark vertex was represented by the Ball-Chiu 
Ansatz which satisfies the relevant symmetries and obeys the Ward-Takahashi
identity and is completely specified by the amplitudes of the dressed quark
propagator.  Only the first term of \Eq{Gammapi} for 
$\Gamma_\pi$ was retained and the chiral limit axial Ward identity result
\mbox{$E_\pi(k;P) \rightarrow B(k^2)/f_\pi$} was used. 

It was verified that the axial anomaly result for the coupling 
constant was obtained (independent of the details of the quark propagator
parameterization used) as a test of the numerical work.  The remaining
pion BS amplitudes from \Eq{Gammapi} evidently do not contribute to the 
chiral limit coupling constant and this has been recently demonstrated
analytically.~\cite{MR98}   However such terms can become increasingly
important at higher mass scales and, for example, are crucial for the
asymptotic behavior of the pion charge form factor.~\cite{MR98}   For this
reason, the large momentum behavior of the calculation~\cite{FMRT95} of 
the axial anomaly transition form factor shown in \Fig{fig:pgg} can be 
expected to receive significant corrections when the sub-leading pion BS
amplitudes are included.   This is a difficult task presently under study.

\section{$\pi NN$ Coupling}

One can ask whether the dynamical content of the simple pion BS amplitude 
of \Eq{sep_pi} produces a $\pi NN$ coupling constant consistent with the
well-established empirical result \mbox{$g_{\pi NN}= 13.4$}.   We make an 
estimate using the valence quark states of a mean field 
chiral quark-meson model~\cite{FT92,B98} of the nucleon in which
the chiral meson modes  are generated as $\bar{q}q$ correlations. This
approach has previously proved fruitful for the $\rho NN$ and $\omega NN$
couplings.~\cite{TQB98}  In Euclidean metric, the $\pi NN$ vertex is 
\beq
\vec{\Lambda}_{\pi NN}(Q)=\case{1}{Z_N}\langle N|\inttp \; 
\bar q(p+\case{Q}{2})\;\vec{\Gamma}_\pi(p;Q)\; q(p-\case{Q}{2})|N \rangle
\label{VpiNN}
\eeq
where $q$ is the quark field, $Q$ is the $\pi$ momentum, $|N \rangle$ is 
the static mean field nucleon state and 
$\vec{\Gamma}_\pi$ is the BS amplitude.  The nucleon valence quark wave 
function renormalization constant $Z_N$ arises from the dynamical  
nature of the quark self-energy.~\cite{FT92} 
At the $\pi$ mass-shell, 
$\Gamma_\pi$ is normalized in the canonical way such
that it is the residue of the pseudoscalar $\bar q q$ propagator there. 

The standard form factor $F_{\pi NN}(Q^2)$ is identified from
recasting the results from Eq.~(\ref{VpiNN}) into the form
\beq
\vec{\Lambda}_{\pi NN}(Q)= \bar u_N(\case{\vec Q}{2})[\;i\gamma_5 
            \vec{\tau}_N F_{\pi NN}(Q^2)\;]u_N(\case{-\vec Q}{2})~.
\label{FpiNN}
\eeq
%-------------------------------------------------------------------
\begin{figure}[htb]

\ihsp \centering{\epsfig{figure=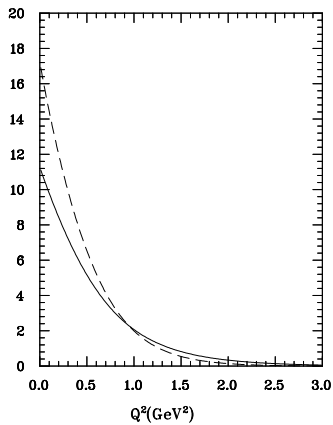,height=5.5cm} } \ihsp
\caption{Pion-nucleon form factor normalized such that $g_{\pi NN}$ is
the value at the pion mass shell value  \mbox{$Q^2 \approx 0$}.  The solid
line includes both PS and PV components of the pion, while the dashed line
includes the PS part only. } 
\label{fig:piNN}

\end{figure}
%-------------------------------------------------------------------
The nucleon mass shell condition does not allow distinction to be made
between pseudoscalar or pseudovector coupling, and only one form factor
can be identified.   Such is not the case for the constituent quarks and
there will be distinct contributions to $F_{\pi NN}$ from the PS and PV
components of the $\pi$ BS amplitude. 
As a mean field nucleon model cannot properly address the recoil issue, we 
resort to the common prescription in which Breit frame kinematics is used
to define a form factor at low momentum transfer.  
  
In Fig.~\ref{fig:piNN} we show the result~\cite{Qian98} for $F_{\pi NN}(Q^2)$
for space-like $\pi$ momentum.  With both PS and PV components of the
pion included we obtain  \mbox{$g_{\pi NN} \approx 11$}, while use of just
the PS pion gives  \mbox{$g_{\pi NN} \approx 17$}.  Rather than compare 
directly to the empirical value $13.4$, the only conclusion we
draw from this estimate is that, without the PV component of the $\pi$ BS 
amplitude, $g_{\pi NN}$ would be overestimated by almost 50\%.   

\section{Summary}
 
Since the parameters in this approach have been previously fixed through the 
requirement that soft chiral quantities such as $m_{\pi/K}$, $f_{\pi/K}$ and charge
radii $r_{\pi/K}$ be reproduced, the meson couplings discussed here
have been produced without adjusting parameters.   The results imply 
that this present approach to modeling QCD for low-energy hadron physics 
can capture the dominant infrared physics. 
The PV component of the pion is found to be important (at the
level of about 25\% and above) for a variety of physical 
quantities such as $m_\pi, f_\pi, g_{\rho\pi\pi}, {\rm and}~ g_{\pi NN}$.  
 We expect that the large 
momentum behavior of form factors such as $\gamma \pi\pi$ and 
$\pi \gamma \gamma $ will require attention to both types of PV amplitude
evident in \Eq{Gammapi}.

\section*{Acknowledgments}
Thanks are due to H. M. Fried and B. Muller for the organization 
of a fine Workshop.  This work is partly supported  by the US National 
Science Foundation under grant No. PHY97-22429.

\end{document}